\documentclass[12pt]{article}
\usepackage{graphicx}
\def \sef{\sin^2 \theta_{\rm eff}}

\begin{document}
\rightline{CLNS 04/1874}
\rightline{hep-ph/0404264}
\bigskip
\centerline{\bf PRECISION ELECTROWEAK TESTS WITH $\bar \nu_e e$
SCATTERING\footnote{To be submitted to Phys.\ Rev.\ D, Brief Reports.}}
\bigskip

\centerline{Jonathan L. Rosner~\footnote{rosner@hep.uchicago.edu.  On leave
from Enrico Fermi Institute and Department of Physics,
University of Chicago, 5640 S. Ellis Avenue, Chicago, IL 60637}}
\centerline{\it Laboratory of Elementary Particle Physics}
\centerline{\it Cornell University, Ithaca, NY 14850}
\medskip

\begin{quote}
Measurements of the cross section for $\bar \nu_e e^-$ elastic scattering
with unprecedented precision have recently been proposed.  The impact of
these experiments for detecting possible deviations from the standard
electroweak theory is analyzed and compared with that of several other
measurements.
\end{quote}

\leftline{PACS Categories:  12.15.Lk, 12.15.Mm, 13.15.+g, 14.60.Lm}
\bigskip

Precise tests of the electroweak theory are able to determine the presence of
``oblique corrections'' affecting vacuum polarization of the photon, $Z$, and
$W$ bosons through new particles in loops.  A language for dealing with these
effects has been developed by Peskin and Takeuchi \cite{Peskin:1990zt} in
terms of two parameters $S$ and $T$, upon which observables depend linearly.
$S = T = 0$ may be defined to correspond to ``no new physics,'' given nominal
values of the top quark and Higgs boson masses $m_t$ and $M_H$.  Both $S$ and
$T$ depend logarithmically on $M_H$, while $T$ depends quadratically on $m_t$.
Constraints on $S$ and $T$ thus can provide information on the mass of the
as-yet-undiscovered Higgs boson as well as restricting the types of new
particles that may enter into gauge boson vacuum polarization loops.

Every new experiment can be analyzed in terms of the constraints it imposes
on $S$ and $T$.  Thus, for example, it was discovered that the weak charge
$Q_W$ measured in parity-violation experiments on heavy atoms such as cesium
\cite{Marciano:1990dp,Sandars} is sensitive almost exclusively to $S$.

Recently a measurement of the total cross section for $\bar \nu_e e^-$ elastic
scattering with unprecedented accuracy has been proposed \cite{Conrad:2004gw}.
In the present note I analyze the potential constraints on $S$ and $T$
following from such a measurement at the proposed 1.3\% level.  The
measurement is found to have much more dependence on $S$ than on $T$, and to
restrict $S$ more closely than measurements of atomic parity violation in
the best-studied cesium \cite{Kuchiev:2003pk} case.  Its impact is compared
with those of several other measurements, including the direct $W$ mass
determination from hadron and $e^+ e^-$ colliders, $M_W = 80.425 \pm
0.034$ GeV/$c^2$ \cite{MW}, and the NuTeV measurement of the ratio of
neutral-current to charged-current cross sections in deeply inelastic neutrino
scattering \cite{Zeller:2001hh}.

The differential cross section for $\bar \nu_e e^- \to \bar \nu_e e^-$ may be
written as a function of recoil electron kinetic energy $T$
\cite{Conrad:2004gw} in the standard electroweak theory as
\begin{equation}\label{eqn:sig}
\frac{d \sigma}{dT} = \frac{G_F^2 m_e}{2 \pi} \left[ (g_V + g_A)^2
+ (g_V - g_A)^2 \left( 1 - \frac{T}{E_\nu} \right)^2 + (g_A^2 - g_V^2)
\frac{m_e T}{E_\nu^2} \right]~~~,
\end{equation}
where $G_F = 1.16639(1) \times 10^{-5}$ GeV$^{-2}$ is the Fermi coupling
constant, $m_e$ is the electron mass, $E_\nu$ is the energy of the incident
$\bar \nu_e$, and the couplings in the lowest-order electroweak theory are
$g_A = -1/2$, $g_V = 1/2 + 2x$, with $x \equiv \sin^2 \theta$, where $\theta$
is the weak mixing angle.  The combination $g_V + g_A = 2x$ is due entirely to
$Z$ exchange, while the combination $g_V - g_A = 1 + 2x$ contains a
contribution of $+2$ from $W$ exchange in the direct channel and $-1 + 2x$ from
$Z$ exchange.  One can then write down the $S$ and $T$ dependence of these
combinations by noting that they become
\begin{equation}\label{eqn:gvga}
g_V + g_A = 2x \rho~,~~ g_V - g_A = 2 + (-1 + 2 x) \rho~,~~ 
g_V = 1 + (2 x - \frac{1}{2}) \rho~,~~g_A = -1 + \frac{\rho}{2}
\end{equation}
when oblique corrections are included, where
\cite{Peskin:1990zt,Marciano:1990dp,Rosner:2001ck}
\begin{equation}\label{eqn:STdep}
x = x_0 + 0.0036 S - 0.0026 T~~,~~~
\rho = 1 + \alpha T = 1 + 0.0078 T~~,
\end{equation}
where $x_0$ is the nominal value of $\sin^2 \theta$ at $S=T=0$.  The parameter
$\sin^2 \theta$ in this discussion is to be interpreted as $\sef$, the
effective value of $\sin^2 \theta$ as measured via leptonic vector and
axial-vector couplings: $\sef \equiv (1/4)(1 - [g_V^{\ell}/g_A^{\ell}])$.
Its latest value in one analysis \cite{Altarelli:2004fq} is
$\sef = 0.23150 \pm 0.00016$.  One can then substitute Eq.\ (\ref{eqn:STdep})
into Eq.\ (\ref{eqn:gvga}) and linearize in $S$ and $T$ to find
\begin{equation}\label{eqn:gST}
g_V = 1/2 + 2 x_0 + 0.0073 S - 0.0055 T~~,~~~g_A = -1/2 + 0.0039 T
\end{equation}

\begin{figure}
\begin{center}
\includegraphics[height=4.65in]{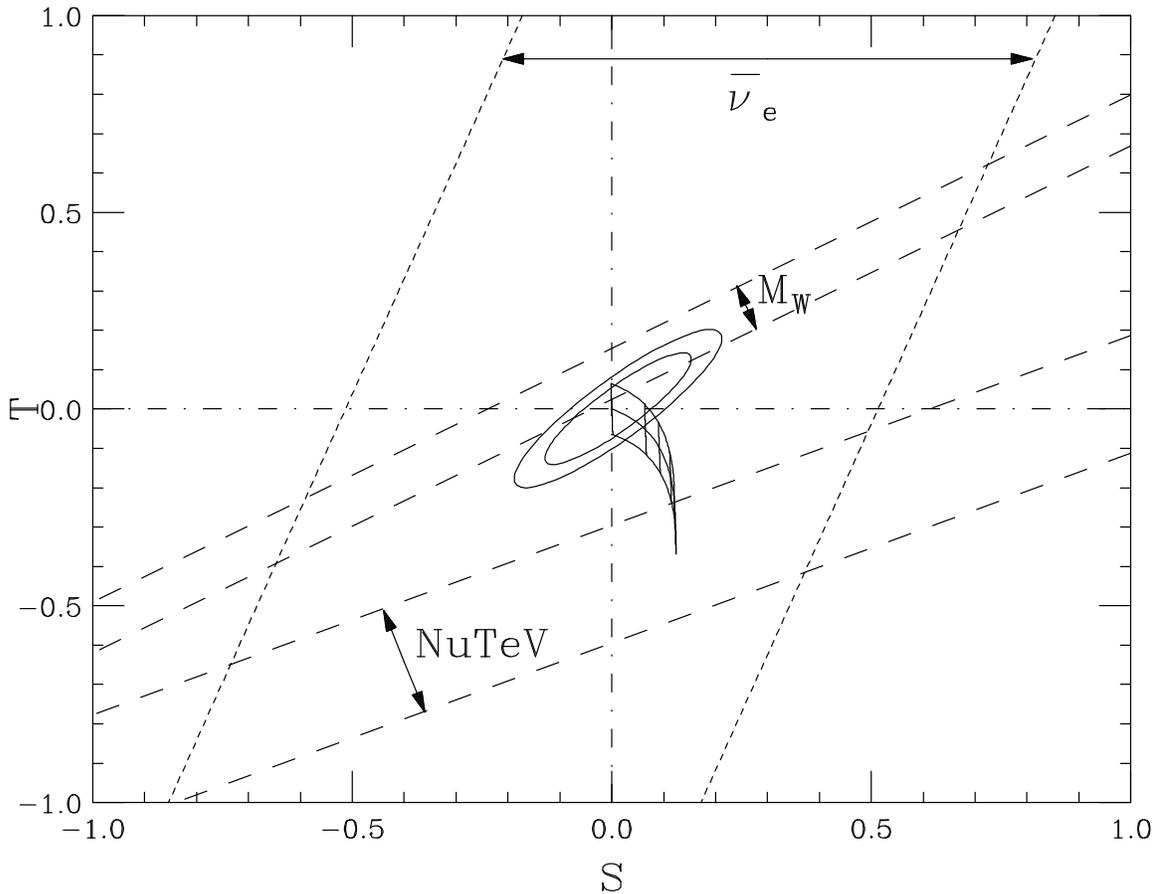}
\caption{Regions of 68\% (inner ellipse) and 90\% (outer ellipse) confidence
level values of $S$ and $T$ based on comparison of theoretical and
experimental electroweak observables \cite{Rosner:2001ck}.
Dash-dotted lines denote the axes $S=0$ and $T=0$.  Diagonal long-dashed lines
denote the constraints from $M_W$ \cite{MW} (above the ellipses) and NuTeV
\cite{Zeller:2001hh} (below the ellipses).  Diagonal short-dashed lines
denote the constraints from the proposed measurement of $\sigma(\bar \nu_e
e^- \to \bar \nu_e e^-)$, assuming a central value entailing $S=T=0$.
Curves emerging from the center of the ellipses denote Standard Model
predictions.  Nearly vertical lines correspond, from left to right, to Higgs
boson masses $M_H = 100,$ 200, 300, 500, 1000 GeV; drooping curves correspond,
from top to bottom, to $+1 \sigma$, central, and $-1 \sigma$ values of $m_t$.
\label{fig:STnue}}
\end{center}
\end{figure}

The number of expected events in the proposal of Ref.\ \cite{Conrad:2004gw}
depends on the coupling constants in the following manner \cite{MS}:
\begin{equation}\label{eqn:nevg}
N = 45950 (g_V + g_A)^2 + 2277 (g_V - g_A)^2 + 4424 (g_A^2 - g_V^2)~~~,
\end{equation}
where the large disparity between the first two coefficients arises from the
fact that the experiment tends to be sensitive to high electron recoil
energies, for which the second term in Eq.\ (\ref{eqn:sig}) is small.
Taking the expressions (\ref{eqn:gST}) for the couplings, one then finds
\begin{equation}\label{eqn:nevST}
N = 11727 + 297 S - 101 T~~~.
\end{equation}
If $N$ is measured to $\pm 1.3\%$, and if a central value consistent with
$S=T=0$ is found, a band $\pm 152 = 297 S - 101 T$, or
\begin{equation}\label{eqn:STband}
\pm 1 = 1.95 S - 0.66 T
\end{equation}
is found.  The results of this constraint are compared with several others in
Fig.\ \ref{fig:STnue}.  The ellipses are based on a previous fit
\cite{Rosner:2001ck} to electroweak data, which have not changed greatly
subsequently.

To put the constraints from $\sigma(\bar \nu_e e^- \to \bar \nu_e e^-)$ in
perspective, no other electroweak observable aside from atomic parity
violation has such a large ratio of $S$ to $T$ dependence.  For comparison,
the latest determination of the weak charge $Q_W$ in cesium finds
\cite{Kuchiev:2003pk} $Q_W({\rm Cs}) = -72.84 \pm 0.49$, to be compared with
the Standard Model prediction \cite{Marciano:1990dp,Takeuchi:1999ci}
$Q_W({\rm Cs}) = -(73.19 \pm 0.13) - 0.800S -0.007T$, thus entailing $S =
-0.45 \pm 0.61$, a band so wide that it cannot be fully displayed in Fig.\
\ref{fig:STnue}.

Thus, though a measurement of $\sigma(\bar \nu_e e^- \to \bar \nu_e e^-)$
at the percent level is not likely to restrict the ellipses in precision
electroweak fits to $S$ and $T$, it provides unique information in much
the same spirit as atomic parity violation at levels superior to those
currently obtained.

I thank J. Conrad and M. Shaevitz for helpful comments, and M. Tigner for
extending the hospitality of the Laboratory for
Elementary-Particle Physics at Cornell during this research.  This work was
supported in part by the United States Department of Energy through Grant No.\
DE FG02 90ER40560 and in part by the John Simon Guggenheim Memorial
Foundation.

% Journal and other miscellaneous abbreviations for references
\def \ajp#1#2#3{Am.~J.~Phys.~{\bf#1}, #2 (#3)}
\def \apny#1#2#3{Ann.~Phys.~(N.Y.) {\bf#1}, #2 (#3)}
\def \app#1#2#3{Acta Phys.~Polonica {\bf#1}, #2 (#3)}
\def \arnps#1#2#3{Ann.~Rev.~Nucl.~Part.~Sci.~{\bf#1}, #2 (#3)}
\def \cmts#1#2#3{Comments on Nucl.~Part.~Phys.~{\bf#1}, #2 (#3)}
\def \cn{Collaboration}
\def \cp89{{\it CP Violation,} edited by C. Jarlskog (World Scientific,
Singapore, 1989)}
\def \dpfa{{\it The Albuquerque Meeting: DPF 94} (Division of Particles and
Fields Meeting, American Physical Society, Albuquerque, NM, Aug.~2--6, 1994),
ed. by S. Seidel (World Scientific, River Edge, NJ, 1995)}
\def \dpff{{\it The Fermilab Meeting: DPF 92} (Division of Particles and Fields
Meeting, American Physical Society, Batavia, IL., Nov.~11--14, 1992), ed. by
C. H. Albright \ite~(World Scientific, Singapore, 1993)}
\def \efi{Enrico Fermi Institute Report No. EFI}
\def \epjc#1#2#3{Eur.~Phys.~J.~C~{\bf #1}, #2 (#3)}
\def \epl#1#2#3{Europhys.~Lett.~{\bf #1}, #2 (#3)}
\def \f79{{\it Proceedings of the 1979 International Symposium on Lepton and
Photon Interactions at High Energies,} Fermilab, August 23-29, 1979, ed. by
T. B. W. Kirk and H. D. I. Abarbanel (Fermi National Accelerator Laboratory,
Batavia, IL, 1979}
\def \hb87{{\it Proceeding of the 1987 International Symposium on Lepton and
Photon Interactions at High Energies,} Hamburg, 1987, ed. by W. Bartel
and R. R\"uckl (Nucl.~Phys.~B, Proc.~Suppl., vol. 3) (North-Holland,
Amsterdam, 1988)}
\def \ib{{\it ibid.}~}
\def \ibj#1#2#3{~{\bf#1}, #2 (#3)}
\def \ichep72{{\it Proceedings of the XVI International Conference on High
Energy Physics}, Chicago and Batavia, Illinois, Sept. 6 -- 13, 1972,
edited by J. D. Jackson, A. Roberts, and R. Donaldson (Fermilab, Batavia,
IL, 1972)}
\def \ijmpa#1#2#3{Int.~J.~Mod.~Phys.~A {\bf#1}, #2 (#3)}
\def \ite{{\it et al.}}
\def \jpb#1#2#3{J.~Phys.~B {\bf#1}, #2 (#3)}
\def \lkl87{{\it Selected Topics in Electroweak Interactions} (Proceedings of
the Second Lake Louise Institute on New Frontiers in Particle Physics, 15 --
21 February, 1987), edited by J. M. Cameron \ite~(World Scientific, Singapore,
1987)}
\def \ky85{{\it Proceedings of the International Symposium on Lepton and
Photon Interactions at High Energy,} Kyoto, Aug.~19-24, 1985, edited by M.
Konuma and K. Takahashi (Kyoto Univ., Kyoto, 1985)}
\def \mpla#1#2#3{Mod.~Phys.~Lett.~A {\bf#1}, #2 (#3)}
\def \nc#1#2#3{Nuovo Cim.~{\bf#1}, #2 (#3)}
\def \np#1#2#3{Nucl.~Phys.~{\bf#1}, #2 (#3)}
\def \pisma#1#2#3#4{Pis'ma Zh.~Eksp.~Teor.~Fiz.~{\bf#1}, #2 (#3) [JETP Lett.
{\bf#1}, #4 (#3)]}
\def \pl#1#2#3{Phys.~Lett.~{\bf#1}, #2 (#3)}
\def \pla#1#2#3{Phys.~Lett.~A {\bf#1}, #2 (#3)}
\def \plb#1#2#3{Phys.~Lett.~B {\bf#1}, #2 (#3)}
\def \pr#1#2#3{Phys.~Rev.~{\bf#1}, #2 (#3)}
\def \pra#1#2#3{Phys.~Rev.~A {\bf#1}, #2 (#3)}
\def \prc#1#2#3{Phys.~Rev.~C {\bf#1}, #2 (#3)}
\def \prd#1#2#3{Phys.~Rev.~D {\bf#1}, #2 (#3)}
\def \prl#1#2#3{Phys.~Rev.~Lett.~{\bf#1}, #2 (#3)}
\def \prp#1#2#3{Phys.~Rep.~{\bf#1}, #2 (#3)}
\def \ptp#1#2#3{Prog.~Theor.~Phys.~{\bf#1}, #2 (#3)}
\def \ptps#1#2#3{Prog.~Theor.~Phys.~Suppl.~{\bf#1}, #2 (#3)}
\def \rmp#1#2#3{Rev.~Mod.~Phys.~{\bf#1}, #2 (#3)}
\def \sci#1#2#3{Science {\bf#1}, #2 (#3)}
\def \si90{25th International Conference on High Energy Physics, Singapore,
Aug. 2-8, 1990}
\def \slc87{{\it Proceedings of the Salt Lake City Meeting} (Division of
Particles and Fields, American Physical Society, Salt Lake City, Utah, 1987),
ed. by C. DeTar and J. S. Ball (World Scientific, Singapore, 1987)}
\def \slac89{{\it Proceedings of the XIVth International Symposium on
Lepton and Photon Interactions,} Stanford, California, 1989, edited by M.
Riordan (World Scientific, Singapore, 1990)}
\def \smass82{{\it Proceedings of the 1982 DPF Summer Study on Elementary
Particl)e Physics and Future Facilities}, Snowmass, Colorado, edited by R.
Donaldson, R. Gustafson, and F. Paige (World Scientific, Singapore, 1982)}
\def \smass90{{\it Research Directions for the Decade} (Proceedings of the
1990 Summer Study on High Energy Physics, June 25--July 13, Snowmass, Colorado),
edited by E. L. Berger (World Scientific, Singapore, 1992)}
\def \tasi90{{\it Testing the Standard Model} (Proceedings of the 1990
Theoretical Advanced Study Institute in Elementary Particle Physics, Boulder,
Colorado, 3--27 June, 1990), edited by M. Cveti\v{c} and P. Langacker
(World Scientific, Singapore, 1991)}
\def \yaf#1#2#3#4{Yad.~Fiz.~{\bf#1}, #2 (#3) [Sov. J. Nucl. Phys. {\bf #1},
#4 (#3)]}
\def \zhetf#1#2#3#4#5#6{Zh.~Eksp.~Teor.~Fiz.~{\bf #1}, #2 (#3) [Sov. Phys. -
JETP {\bf #4}, #5 (#6)]}
\def \zpc#1#2#3{Zeit.~Phys.~C {\bf#1}, #2 (#3)}
\def \zpd#1#2#3{Zeit.~Phys.~D {\bf#1}, #2 (#3)}

\end{document}